\documentstyle[preprint,aps]{revtex}
\begin{document}
\draft
\preprint{Submitted to Phys. Lett. {\bf A}}
\title{
Inhomogeneity and Current Corrections to a Non-uniform Electronic System in
Strong Magnetic Fields
}
\author{
A. Mazzolo$^{1,2}$, E. L. Pollock$^{3}$ and G. Z\'erah$^{2}$
}
\address{
$^1$Laboratoire de Physique Th\'eorique et Hautes Energies,
91405 Orsay Cedex, France
\\$^2$Centre d'Etudes de Limeil-Valenton, 94195 Villeneuve St. Georges
Cedex, France
\\$^3$Physics Department, Lawrence Livermore National Laboratory,
University of California, Livermore, California 94550
}

\date{\today}

\maketitle

\begin{abstract}

Using the polarizability of a free electron gas in a magnetic
field and the Current-Density Functional Theory (CDFT)
developed by Vignale and Rasolt, we derive the gradient and current
corrections for the energy functional of a non-uniform electronic system
in a strong magnetic field.

First, we find the Tomishima-Shinjo functional by neglecting the current
variation. Taking into account the current variation leads to new
gradient terms which change contraction effects in the direction
perpendicular to the magnetic field.
\end{abstract}


\section{Introduction}
In a magnetic field the electronic density is elongated along the field
direction. There are many intuitive ways of seeing this. For example, in a
path integral description, since the zero field propagator is multiplied by
the phase factor $exp \bigl[ {ie \over {{\hbar}c}}\oint{\bf A}\cdot d\bf
l\bigr]$, each closed path is reweighted by a phase proportional
to the magnetic flux it encloses. In the average over paths which produces
the electronic density, paths parallel to the field, enclosing no flux,
receive a weight of one; while paths perpendicular to the field, enclosing
flux, receive a reduced weight. The resulting average for the electronic
density, thus, has an elongated shape.\

The lowest order statistical model however leads to an isotropic
density[1]. The anisotropy requires gradient corrections [2]. The necessity of
going beyond the lowest order has also spawned a variety of approximate
methods to regain anisotropy [3,4,5,6,7,8].\

In this Letter, we first review the fundamental results of  CDFT.
Then, neglecting the current variation, we rederive the gradient corrections
of Tomishima and Shinjo. Finally, taking this variation into account, we
calculate all second order electronic density gradient corrections.

\section{Current density functional formalism}

For systems in a magnetic field, Vignale and Rasolt [9]
show that the density $\rho$ and the paramagnetic current density
${\bf j}_p$, related to the gauge-invariant total current density, by
\begin{eqnarray}
        {\bf j({\bf r})} = {\bf j_p({\bf r})} +{e\over
	mc}\rho({\bf r})A({\bf r})
\label{current}
\end{eqnarray}
uniquely determine the external scalar, $V_{x}({\bf r})$, and vector
potential $\bf A({\bf r})$.
The ground state energy functional, which may be written as
\begin{eqnarray}
 & &E[\rho,{\bf j}_p] = K[\rho,{\bf j}_p] +\int d{\bf r} \, \rho({\bf r}) V_{x}
({\bf r}) +  {e^2\over2} \int \int d{\bf r}\, d{\bf r}' {{\rho({\bf
r})\rho({\bf r}')}\over{|{\bf r}-{\bf r}'|}} \nonumber\\
 & & \mbox{\ \ \ \ \ \ \ \ \ \ \ \ \ } + {e \over c}\int d{\bf r} \,
{\bf j}_p({\bf r}){\bf A}({\bf r}) + {e^2\over 2mc^2}\int d{\bf r} \,
\rho({\bf r}){\bf A^2({\bf r})} + E_{xc}[\rho,{\bf j}_p]
\label{energie}
\end{eqnarray}
where $K[\rho,{\bf j}_p]$ is the kinetic energy of a free electron gas
in the external potentials $V_{x}({\bf
r})$ and $\bf A({\bf r})$, and $E_{xc}[\rho,{\bf j}_p]$ is the
exchange-correlation energy, is minimized at the ground state energy
by the correct density and current distribution.
In the spirit of Thomas-Fermi approximations we subsequently ignore the
exchange-correlation term and the Hartree mean field term.

Although $K[\rho,{\bf j}_p]$ is not invariant under a gauge transformation:
\begin{eqnarray}
 & &\bf A({\bf r})\to \bf A({\bf r})-\bf \nabla\chi \nonumber\\
 & &\bf j_p({\bf r})\to \bf j_p({\bf r})+ {e\over mc}\rho({\bf r}) \bf
\nabla\chi
\label{gauge}
\end{eqnarray}
the difference:
\begin{eqnarray}
U[\rho,{\bf j}_p] = K[\rho,{\bf j}_p]-
{m\over2}\int d{\bf r} \, {{\bf j_p\bf
j_p}\over\rho}=E[\rho,{\bf j}_p]-\int d{\bf r} \, \rho({\bf r}) V_{x}({\bf r})
-{m\over2}\int d{\bf r} \, {{\bf j\bf j}\over\rho}
\label{Uen}
\end{eqnarray}
which depends on the energy and physical current, is gauge invariant
and, since unchanged by the addition of the gradient of an arbitrary function,
depends on ${\bf j}_p$  only through
\mbox{$\bf \nu({\bf r})=\bf \nabla\times \bigl(\bf j_p /\rho\bigr)$},
the vorticity introduced by Vignale and Rasolt [9].

     In the next two sections we determine the gradient expansion for $U$.

\section{Gradient corrections in strong fields}

We consider an electron gas in an uniform magnetic field ${\bf B}=
B_{0}\hat{{\bf z}}$. As a first step we will neglect the variation of
vorticity. Then the functional U is reduced to a functional of
only one variable, $\rho$.
We postulate an expression for U with gradient
terms:
\begin{eqnarray}
U[\rho] = \int d{\bf r} \, \biggl\{ f(\rho({\bf r}))+g_{\bot}(\rho({\bf
r})){\bigl|} {\bf \nabla}_{\bot}\rho {\bigr|}^2+g_{\|}(\rho({\bf
r})){\bigl|} {\bf \nabla}_{\|}\rho {\bigr|}^2 \biggr\}
\label{U}
\end{eqnarray}
where $\bot$ and $\|$ denote orientations perpendicular and parallel to the
field.

The function $f(\rho({\bf r}))$ can be determined from the Euler equation
\begin{eqnarray}
\begin{array}{c}\underline{\delta E[\rho({\bf r})]}\\\delta\rho({\bf r})
\end{array}=\mu\;.
\end{eqnarray}
For the assumed $U$ this gives
\begin{eqnarray}
 & &f'(\rho({\bf r})) -2g_{\bot}(\rho({\bf
r})) {\bf \nabla}_{\bot}^2\rho -2g_{\|}(\rho({\bf
r})) {\bf \nabla}_{\|}^2\rho \nonumber\\
 & & \mbox{\ \ \ \ } - g_{\bot}'(\rho({\bf
r})){\bigl|} {\bf \nabla}_{\bot}\rho {\bigr|}^2 -g_{\|}'(\rho({\bf
r})){\bigl|} {\bf \nabla}_{\|}\rho {\bigr|}^2=-V_{x}({\bf r})+\mu
\label{equa}
\end{eqnarray}
which, at uniform density  $\rho_0$ and no external potential, is simplified
to:
\begin{eqnarray}
       f'[\rho_0]=\mu_0
\label{equa1}
\end{eqnarray}
where $\mu_0$ is the chemical potential of a free electron gas.

$g_{\bot}(\rho({\bf r}))$ and $g_{\|}(\rho({\bf r}))$  will be determined by
comparing the calculated linear density response  to a variation in the
external potential, $\delta\rho(\bf q)=\chi(\bf q)\delta{V_x}(\bf q)$, to
the free electron gas results. A second variation of the Euler
equation gives (after a Fourier transform),
\begin{eqnarray}
f''(\rho_0) +2g_{\bot}(\rho_0){\bf k}_{\bot}^2
	    +2g_{\|}(\rho_0){k}_{\|}^2=-1/\chi(\bf k)
\label{equa3}
\end{eqnarray}
at uniform density  $\rho_0$.

      As a preliminary example recall the case of no magnetic field.
The chemical potential $\mu=\hbar^{2}k_{F}^{2}/2m=
\hbar^{2}(3\pi^{2}\rho)^{2/3}/2m$  so eqn.~(\ref{equa1}) reproduces
the usual Thomas-Fermi result,
\begin{eqnarray}
f(\rho)={3\over 10}{\hbar^{2}\over m}(3\pi^{2})^{2/3}\rho^{5/3}
\end{eqnarray}
and the Lindhard response formula
\begin{eqnarray}
\chi( k)=-{m k_{F}\over \hbar^{2} \pi^{2}}\left[ {1\over 2}+
\begin{array}{c}\underline{ 2( k_{F}^{2}- k^{2})}\\ k_{F} k
\end{array}
\ln\left|
\begin{array}{c}\underline{2 k_{F}- k}\\2 k_{F}+ k\end{array}
 \right| \right]
=-{m k_{F}\over \hbar^{2} \pi^{2}}\left(1-{1\over 3}( k/2 k_{F})^{2}\right)
+O(k^{4})
\end{eqnarray}
or
\begin{eqnarray}
\chi^{-1}( k)=-{\hbar^{2}\pi^{2}\over m  k_{F}}\left(1+{1\over 3}(  k/
2 k_{F})^{2} \right)+ O( k^{4})\;.
\end{eqnarray}
gives, on comparison with eqn.~(\ref{equa3}), the von Weizsacker gradient
corrections
\begin{eqnarray}
g_{\bot}=g_{\|}={\hbar^{2} \over 3 m}{\pi^{2}\over 8 k_{F}^{3}}=
{1\over 9}{1\over 8 \rho}{\hbar^{2}\over m}
\end{eqnarray}
(with the $1/9$ factor from the long wavelength comparison)[10].

In the case of
very strong fields, where all the electrons are in the lowest Landau
level and spin polarized, the free electron gas chemical potential
$\mu_0=\hbar\omega_{c}/2+ \hbar^{2} k_{F}^{2}/2m$ with
$k_{F}=2\pi^{2}\hbar\rho/m\omega_{c}$ so

\begin{eqnarray}
        f(\rho)={{\hbar\omega_c} \over 2}\rho+{{2\pi^4\hbar^4}\over
{3m^3{\omega_c}^2}}{\rho}^3
\label{equa2}
\end{eqnarray}
The free electron gas susceptibility for this case is
well known since Horing [11,12]
\begin{eqnarray}
\chi({\bf k})=-{2\over {4{\pi^2}{\omega_c}l^4 \hbar}}
      {{F({\bf k}_{\bot})} \over{{k}_{\|
}}} \mbox{\ ln \ } \Biggl| {{2k_F+{k_\|}} \over {2k_F-{k_\|}}} \Biggr|
\label{equa4}
\end{eqnarray}
where ${F({\bf k})}=e^{- k^2 l^2/2}$ and $l$ is the
associated magnetic length defined by: ${l^2}=\hbar c/eB_0$.
The expansion of $\chi^{-1}$ to second order in $\bf k$
gives:
\begin{eqnarray}
  -{\chi^{-1}(\bf k)}={{4{\pi^4}{\hbar^4}\rho}\over {m^3{\omega_c}^2}}+
{{2{\pi^4}{ \hbar^5}\rho}\over {m^4{\omega_c}^3}}{{\bf k}_{\bot}^2}
-{{\hbar^2} \over {12\rho m}}{{k_\|}^2}
\label{equa5}
\end{eqnarray}
so
\begin{eqnarray}
g_{\bot}={\pi^{4}\hbar^{5}\rho\over m^{4}\omega_{c}^{3}}
\end{eqnarray}
and

\begin{eqnarray}
g_{\|}= -{1\over 24 \rho}{\hbar^{2}\over m}
\end{eqnarray}

The resulting energy functional:

\begin{eqnarray}
E[\rho] = \int d{\bf r} \, \biggl\{ {{{\hbar\omega_c} \over
2}}\rho+{{{2\pi^4\hbar^4}\over
{3m^3{\omega_c}^2}}{\rho}^3}+{{{\pi^4}{\hbar^5}}\over
{m^4{\omega_c}^3}}{\rho{\bigl|} {\bf \nabla}_{\bot}\rho {\bigr|}^2}-{{\hbar^2}
\over 24m}{{\bigl|}{{\bf \nabla}_{\|}\rho {\bigr|}^2} \over \rho}
 \biggr\} +\int d{\bf r} \, \rho V_{x}
\label{equa6}
\end{eqnarray}
is the same as the one originally derived by Tomishima and Shinjo [6].\\

Note that even though the free electron states are still plane waves along
the field direction the gradient correction for this direction
has the opposite sign from the free field case. This is a consequence of
the 1-dimensional nature of the high field limit. Explicitly, the free
fermion response function in 1-dimension is

\begin{eqnarray}
\chi(k) ={-m\over \hbar^{2} \pi k}
 \mbox{\ ln \ } \Biggl| {{2k_F+{k}} \over {2k_F-{k}}} \Biggr|
={-m\over \hbar^{2}\pi k_{F}}\left\{1+{1\over
12}k^{2}/k_{F}^{2}\right\}+O(k^{4})
\end{eqnarray}
or
\begin{eqnarray}
\chi^{-1}(k)={-\hbar^{2}\pi k_{F}\over m}\left\{1-{1\over
12}k^{2}/k_{F}^{2}\right\}
   +O(k^{4})
\end{eqnarray}
with $k_{F}=\pi\rho$. Therefore [13]

\begin{eqnarray}
g_{\|}=-{\pi\over 24 k_{F}}{\hbar^{2}\over m}=-{1\over 24\rho}{\hbar^{2}\over
m}
\end{eqnarray}
which is exactly the high magnetic field result derived above.

\section{Current corrections}
The Tomishima and Shinjo approximation is obtained
when  the variation of vorticity is omitted.
However, this variation involves the gradient of
$\rho$ and can not a priori be neglected. We therefore generalize our
previous expression for $U$ to include vorticity as follows:

\begin{eqnarray}
U[{\rho},{\bf \nu}] = \int d{\bf r} \, \biggl\{ f(\rho({\bf r}),{\bf
\nu({\bf r})})+g_{\bot}(\rho({\bf r}),{\bf \nu({\bf r})})
{\bigl|} {\bf \nabla}_{\bot}\rho {\bigr|}^2+g_{\|}(\rho({\bf r}),{\bf
\nu({\bf r})})
{\bigl|} {\bf \nabla}_{\|}\rho {\bigr|}^2 \biggr\}
\label{Urho}
\end{eqnarray}
{}From the definition of the vorticity:

\begin{eqnarray}
{\bf \nu({\bf r})}=-{{{eB_0}\over mc}{\bf u}_z}+\bf \nabla\times \bigl({\bf j
\over \rho}\bigr)  \;.
\label{vort}
\end{eqnarray}
Consequently, we do not need to add in the $U[{\rho},{\bf
\nu}]$ expansion terms in
${\bigl|} {\bf \nabla\nu} {\bigr|}$ which are  higher order.

Variation of the energy with respect to ${\bf j}_p$,
 $\delta E/\delta {\bf j}_p=0$, using the chain rule result
\[
{\delta \over \delta {\bf j}_{p}({\bf r})}
={1\over{\rho({\bf r})}}
{\bf \nabla}\times {\delta\over{\delta\nu ( {\bf r})}}
\]
gives:
\begin{eqnarray}
{\bf j}=-{1 \over m}{\bf \nabla\times} {{\partial f}\over{\partial{\bf
\nu}}}
\label{cou5}
\end{eqnarray}
to second order.
The Euler equation, $\delta E/\delta\rho({\bf r})=\mu$, becomes:

\begin{eqnarray}
 & &{{\partial f}\over{\partial\rho}}[{\rho},{\nu}]
 -2g_{\bot} {\bf \nabla}_{\bot}^2\rho -2g_{\|}
 {\bf \nabla}_{\|}^2\rho \nonumber\\
 & & \mbox{\ } -{{\partial g_{\bot}}\over {\partial\rho}}
{\bigl|} {\bf \nabla}_{\bot}\rho {\bigr|}^2 -{{\partial g_{\|}}\over
{\partial\rho}}
{\bigl|} {\bf \nabla}_{\|}\rho {\bigr|}^2=-V_{x}({\bf r})+\mu+{m\over2}
{{\bf j\bf j}\over\rho^2}
\label{equa15}
\end{eqnarray}
At uniform density we get:$${{\partial
f}\over{\partial\rho}}[{\rho_0},{\nu_0}]=\mu[{\rho_0},{\nu_0}] \mbox{\ \
\ \ with \ \ \ \ } {{\bf \nu}_0}=\omega_{c}\hat{{\bf z}} $$
so the expression for $f$ is formally the same as (\ref{equa2}) with the
 variables $\rho$ and $\bf \nu$:
\begin{eqnarray}
        f[{\rho},{\bf \nu}]={{\hbar \over 2}|{\bf \nu}|}\rho+{{2\pi^4\hbar^4}
\over {3m^3{|{\bf \nu}|}^2}}{\rho}^3 \;.
\label{equa9}
\end{eqnarray}
With equation (\ref{cou5}), we can now calculate the total current density,

\begin{eqnarray}
{\bf j} {\simeq}- \biggl\{ {{\hbar \over 2m}-{{4\pi^4\hbar^4}\over{m^4}}{\rho^2
\over{|\nu_0|^3}}} \biggr\}{{\bf \nabla}\rho\times {\bf u}_z}
\label{equa10}
\end{eqnarray}
where we have replaced $|\bf \nu|$ by $|\bf \nu_0|$ in the second member
of (\ref{equa10}) since the $\bf \nu$ variations are  higher order.
 In the limit
of very strong fields we find the result of Skudlarski and Vignale~[14].
On the other hand equation (\ref{equa3}) remains unchanged at $\rho_0$
since all
the additional terms are proportional to ${\bf \nabla}\rho$. Thus, the
expressions for  $g_{\bot}$, $g_{\|}$ are absolutely the same as in section III
with ${\omega_c} \to {|\bf \nu|}$.

 In return, E is modified by the
presence of the current, and for a comparison with Tomishima and Shinjo
we have to develop $\bf \nu$, eqn. (\ref{vort}), using eqn. (\ref{equa10}):

\begin{eqnarray}
    {|{\bf \nu}|} = {\omega_c} - {\hbar \over 2m}{\biggl[}{{{\bigl|}{{\bf
\nabla}_{\bot} \rho{\bigr|}^2} \over \rho^2} - {{{\bf
\nabla}_{\bot}^2 \rho} \over \rho}}{\biggr]} - {{{4\pi^4}{\hbar^4}}\over
{m^4{\omega_c}^3}}{\biggl[}{{\bigl|}{\bf
\nabla}_{\bot} \rho{\bigr|}^2}+{\rho{\bf
\nabla}_{\bot}^2 \rho} {\biggr]}   \;.
\label{equa12}
\end{eqnarray}
Using this expansion in eqn. (\ref{equa9}) for $f[\rho,\nu]$ and with
eqn. (\ref{equa10}) for  the current we finally obtain for the energy
functional E:

\begin{eqnarray}
&  E[\rho] = \int d{\bf r} \, &\biggl\{ {{{\hbar\omega_c} \over
2}}\rho+{{{2\pi^4\hbar^4}\over
{3m^3{\omega_c}^2}}{\rho}^3}-{{\hbar^2}\over 8m}{{\bigl|}{{\bf
\nabla}_{\bot}\rho
{\bigr|}^2} \over \rho}+{{{3\pi^4}{\hbar^5}}\over
{m^4{\omega_c}^3}}{\rho{\bigl|} {\bf \nabla}_{\bot}\rho
{\bigr|}^2}\nonumber \\
&  &-8m\left({{{\pi^4}{\hbar^4}}\over{m^4{\omega_c}^3}}\right)^{2}
\rho^{3}\left|{\bf \nabla}_{\bot}\rho\right|^{2}
 -{{\hbar^2}\over 24m}
{{\bigl|}{{\bf \nabla}_{\|}\rho {\bigr|}^2} \over \rho} \biggr\} +\int
d{\bf r} \, \rho V_{x}
\label{equa13}
\end{eqnarray}

\section{Conclusion}

Current corrections do not change anything along the field direction
since this correction is only due of the 1-dimensional nature of the
high field limit. The transverse corrections however, have greatly changed.
First we note an important additional
term in $|{\bf \nabla}_{\bot}\rho |^{2}/  \rho$ which does not
depend on the strong field intensity. Perpendicular gradients are
favored by this term. The next term in ${{\bf
\nabla}_{\bot}\rho}$, which is small compared to the constant term,
is the same as found by Tomishima and Shinjo but with a factor $+3$.
 This term and the third term in
${{\bf \nabla}_{\bot}\rho}$ decrease with $B_0$. Therefore,
contrary to Tomishima and Shinjo the total transverse corrections increase with
$B_0$. This result modifies the transverse pinch.

In summary, as expected, the
behavior of the total corrections produces an anisotropic
density profile. This anisotropy is caused by a reduction of dimensionality
due to the high field limit. Our approach, which is technically simple, allows
us
to extend the calculus to finite temperatures and to include some Landau
levels. But since the minimization of  eqn.~(\ref{equa13}) is rather
complicated,
we leave its numerical solution for a forthcoming article.

\acknowledgements

      	One of us, A.M. would like to thank Dr D. Levesque for his kind
hospitality at LPTHE and the CEA for financial support. LPTHE is Unit\'e de
Recherche de L'Universit\'e Paris XI associ\'ee au CNRS.
Work done at the Lawrence Livermore National Laboratory is supported by the
U.S. Department of Energy under Contract W-4705-Eng-48.

\end{document}